\documentclass[prd,twocolumn,superscriptaddress,showpacs,preprintnumbers,amsmath,amssymb,APS]{revtex4}
\usepackage{bm}
\usepackage[latin1]{inputenc}
\usepackage[spanish,english]{babel}
\usepackage{amsfonts}
\usepackage{amssymb}
\usepackage{graphicx}
\usepackage{color}

\newcommand{\be}{\begin{equation}}
\newcommand{\ee}{\end{equation}}
\newcommand{\bea}{\begin{eqnarray}}
\newcommand{\eea}{\end{eqnarray}}

\begin{document}
\title{{\bf Remarks on the renormalization of primordial cosmological perturbations }}

\author{Ivan Agullo}\email{ivan.agullo@uv.es}
\affiliation{ {\footnotesize Institute for Gravitation and the Cosmos, Physics Department, Penn State, University Park,
PA 16802-6300, U.S.A.}}
\author{Jose Navarro-Salas}\email{jnavarro@ific.uv.es}
\affiliation{ {\footnotesize Departamento de Fisica Teorica and
IFIC, Centro Mixto Universidad de Valencia-CSIC.
    Facultad de Fisica, Universidad de Valencia,
        Burjassot-46100, Valencia, Spain, and}}\affiliation{ {\footnotesize Physics Department, University of
Wisconsin-Milwaukee, P.O.Box 413, Milwaukee, WI 53201 USA}}

\author{Gonzalo J. Olmo}\email{gonzalo.olmo@csic.es }
\affiliation{\footnotesize Departamento de Fisica Teorica and IFIC, Centro Mixto Universidad de Valencia-CSIC.
    Facultad de Fisica, Universidad de Valencia, Burjassot-46100, Valencia, Spain }
\author{Leonard Parker}\email{leonard@uwm.edu}
\affiliation{ {\footnotesize Physics Department, University of
Wisconsin-Milwaukee, P.O.Box 413, Milwaukee, WI 53201 USA}}

\date{August 3, 2011}

\begin{abstract}
We briefly review  the need to perform renormalization of inflationary perturbations to properly work out the physical power spectra. We also summarize the basis of  (momentum-space) renormalization in curved spacetime and address several misconceptions found in recent literature on this subject.

\end{abstract}


\maketitle

\section{Introduction}

The inflationary universe \cite{guth} opens an exciting window to observationally test fundamental aspects of the theory of quantized fields in curved spacetime \cite{parker66, parker69, parker-toms, birrell-davies}. In a curved background, the vacuum energy of  quantum fields enters into the gravitational field equations. As is well known, infinities arise in the  computation of vacuum energy, and other expectation values quadratic in fields, due to the ultraviolet behavior of the field theory. Therefore, the potential gravitational effects of the quantum vacuum must be handled with care.  Methods have been developed to define regularization and renormalization procedures to physically account for the effects of vacuum energy in free and interacting quantum field theory in curved spacetime.

In the cosmological scenario, it was shown in the early eighties that vacuum fluctuations can induce  a primordial spectrum of density perturbations during an inflationary expansion. Remarkably, the calculated spectrum satisfactorily accounts for the origin of the cosmic inhomogeneities that we observe in the present universe. Let $\varphi$ represent a generic field describing (scalar or tensor) perturbations during inflation. The quantum fluctuations of $\varphi$ can be quantified by the mean square fluctuation in the vacuum state (see, for instance, \cite{kolb-turner,books})
\be \label{var} \langle  \varphi^2(\vec{x}, t) \rangle= \int d^3k |\varphi_k(t)|^2 \equiv \int_0^{\infty} \frac{dk}{k} \Delta^2_{\varphi}(k, t) \ . \ee
where $\varphi_k(t)$ is defined by the expansion in Fourier modes of a free field operator
\be
\varphi(\vec{x}, t)= \int d^3k (A_{\vec{k}}\varphi_k(t)+ A^{\dagger}_{-\vec{k}}\varphi^*_k(t))e^{i\vec{k}\vec{x}}
\ ,
\label{eq: field in RW}
\ee
where $A_{\vec{k}}$ and $A^{\dagger}_{\vec{k}}$ are creation and annihilation operators, such that $A_{\vec{k}} |0 \rangle= 0$.  In the last equality of Eq. (\ref{var}) we have defined the power spectrum $\Delta^2_{\varphi}(k, t)$, which is the quantity conventionally used in cosmology to quantify the variance (\ref{var}) .  For a single $k$, the power spectrum $\Delta^2_{\varphi}(k, t)$ is well defined.  However, the formal variance  $\langle \varphi^2(\vec{x}, t)  \rangle$  diverges in the ultraviolet.  One could argue that this ultraviolet divergence concerns ultrashort wavelength modes, and it is not going to affect the predictions regarding the finite range of wavelengths that we actually observe in cosmic inhomogeneities. But in view of the fact that the  perturbations are a consequence of quantum field theory in curved spacetime, it was proposed in  \cite{parker07, grg09, prl09, talk, prd10}  that the physical power spectrum should be defined in terms of the renormaliz
 ed mean square
 \be \label{renvarps}
\langle  \varphi^2(\vec{x}, t)  \rangle_{ren}= \int_0^{\infty} \frac{dk}{k}\tilde\Delta^2_{\varphi}(k, t) \
\ .  \ee
They found that, when methods of renormalization in curved space-time are applied, the power spectrum at wavelengths of observational interest may be affected in ways that can be tested by observations
in the not-too-distant future. 

It has been argued  recently that the renormalization procedure used in \cite{parker07, grg09, prl09, talk, prd10} is not well-defined at the time scales at which cosmic inhomogenities are created during inflation,
particularly at the time $t_k$ that each mode crosses the Hubble sphere \cite{durrer09,marozzi11}.  In the present work, we briefly summarize the basis of  renormalization (including its effects on the momentum-space power spectrum) in an expanding universe and address the criticisms raised in \cite{durrer09,marozzi11}. We find that the arguments and conclusions proposed in \cite{durrer09,marozzi11} are in direct conflict with some of the basic principles of renormalization in curved space-time. The last paragraph in sections \ref{section ii} and \ref{section iii} summarize the arguments provided here in relation to the criticism of \cite{durrer09,marozzi11}.
\section{Renormalization in curved space-time}\label{section ii}

The renormalization of expectation values like those corresponding to the mean square fluctuation and the stress tensor of a quantized field in a curved space-time is more involved than in Minkowski space, even for the simplest case of a free field.  This is rooted in the fact that in a general curved space-time there are insufficient isometries to uniquely determine a global vacuum state.  Moreover, the presence of space-time curvature yields new types of divergences not present in Minkowski space.  This is true for free as well as interacting fields.  

It is instructive to briefly consider, in a general curved space-time, an interacting scalar field $\varphi$ with a $\lambda \varphi^4$ interaction term in its Lagrangian.  In flat space-time, one usually renormalizes such an interacting theory by going to momentum-space and adding counter-terms to the Lagrangian that will cancel the regularized UV divergences in the momentum space integrals corresponding to Feynman diagrams.  The latter method is not directly applicable in a general curved space-time because of the absence of a global and generally-covariant momentum-space expansion. However, this problem was overcome for a general smooth curved space-time in \cite{bunch-parker} by making use of a Riemann normal coordinate system (RNC) with its origin at a given space-time point $x'$. Such a coordinate system, based on the system of geodesics leaving the point $x'$, exists in any smooth space-time in a normal neighborhood of $x'$ (i.e., one in which these outgoing geodesics 
 do not intersect). In  a RNC, the metric infinitesimally close to $x'$ is Minkowskian and has vanishing first derivatives with respect to the space and time coordinates of the RNC. In addition, the metric has a well defined expansion in powers of the RNC coordinates $y$, which are defined for each space-time point $x$ in the normal neighborhood of $x'$ by means of the tangent vector at $x'$ of the unique geodesic that connects $x'$ to $x$ and the invariant length  of that geodesic.  The coefficients in this expansion of the metric in powers of $y$ are {\em constants} formed from contractions of powers of the Riemann tensor evaluated at the origin $x'$ of the RNC system.

In \cite{bunch-parker}, Bunch and Parker (BP) defined a local momentum-space ``Fourier" transform of the Feynman propagator $G(x, x')$ based on an RNC with its origin at $x'$.  Working in a {\em general} curved space-time, they used this local momentum-space method to evaluate the Feynman diagrams necessary to renormalize the {\em interacting} scalar field theory to second order in the interaction coupling constant $\lambda$ appearing in the $\lambda \varphi^4$ self-interaction. Dimensional regularization was used to replace the UV infinities that are present in 4-dimensions by covariant well-defined expressions.   For the interacting theory, they introduced into the Lagrangian the minimal set of generally covariant counter-terms necessary to absorb the regularized UV ``infinities'' into the values of the constant coefficients of these counter-terms. This process leaves one with a Lagrangian having terms of the same form as the original counter-terms, but with ``renormalized'
 ' constant coefficients that are assumed to be finite, well-defined, and in principle measurable.  

The terms that involve the curvature at the space-time point $x'$ were shown to drop out of the final result. This step is non-trivial and necessary for the renormalized interacting theory to be covariant in a general curved space-time. This calculation showed that interacting $\lambda \varphi^4$ theory is renormalizable in a general curved space-time to second order in $\lambda$. The reason for using the {\em minimal} set of counter-terms is to alter the original Lagrangian as little as possible, thus avoiding the arbitrary introduction of interactions that are not necessary for renormalization.

They also calculated the leading terms of their local momentum-space expansion of the propagator that remain 
when $\lambda=0$, (i.e., for the non-interacting, free field).  For example, for a minimally coupled {\em free} field, 
$\varphi$, the local momentum-space Fourier transform of $|g(x)|^{-1/4} G(x,x') |g(x')|^{-1/4}\equiv {\bar G}(x,x')$ has an asymptotic expansion for large $k^2$ that is given by \cite{bunch-parker} 
\be 
{\bar G}(k;x') \sim  \frac{1}{k^2 + m^2} + \frac{R(x')}{6(k^2 + m^2)^2} + \cdots \ , 
\label{eq:BP-expansion} 
\ee
where $k$ is the 4-momentum, $m$ is the mass of the field, and the ellipsis ($\cdots$) includes terms that go as the third and fourth powers of $1/(k^2 + m^2)$.  The coefficients of those terms are formed from contractions of Riemann tensors and their derivatives, evaluated at point $x'$, and are given explicitly in Eq. (2.21) of \cite{bunch-parker}.  The Feynman propagator in the normal neighborhood of $x'$ is defined as in Minkowski space by replacing $m^2$ by $m^2 - i\epsilon$, with $\epsilon$ small and positive, and carrying out the $k_0$ integration along the real axis in the complex plane. 

By carrying out the inverse Fourier transformation back to curved space-time, they were able to recast this series into the form of the proper-time or heat kernel expansion of ${\bar G}(x,x')$.  They showed that the coefficients that they had calculated (which contain all terms up to fourth order in derivatives of the metric) in the series in Eq.(\ref{eq:BP-expansion}), when evaluated in the coincidence limit ($x\rightarrow x'$), are identical to the corresponding coefficients of the proper-time series that had been calculated \cite{dewitt75} for the free field in the coincidence limit.

As is well known, the expectation values $\langle T_{\mu \nu} (x') \rangle$ of the energy-momentum tensor of the free scalar field in physical states can be found from the exact expression for the Green function $G(x,x')$ by applying to $G(x,x')$ the appropriate second-order differential operator that produces the expression for the operator $T_{\mu \nu} (x')$ in the coincidence limit as $x\rightarrow x'$ and taking the expectation value in the physical state of interest.  From the asymptotic series for ${\bar G}(k;x')$ discussed above, it is clear that the terms in that series up to those that go as $1/(k^2 + m^2)^3$, will produce divergent contributions to  $\langle T_{\mu \nu} (x') \rangle$ when the Riemann normal coordinate $y$ is taken to zero and the dimension $n$ is taken to the value $4$ in the $n$-dimensional Fourier transform.  It can be seen that, when the two spatial derivatives in $T_{\mu \nu}$ are taken into account, the first term in (\ref{eq:BP-expansion}) giv
 es a contribution to $\langle T_{\mu \nu} (x') \rangle$ that has a quartic UV divergence that does not depend on the curvature at $x'$, the second term has a quadratic UV divergence that depends on the Ricci scalar curvature $R(x')$, and the subsequent terms have logarithmic UV divergences that depend in a more complicated way on contractions and spatial derivatives of the Riemann tensor, evaluated at $x'$.  

It is also clear that these UV divergent contributions to $\langle T_{\mu \nu} (x') \rangle$ have the same state-independent expression in terms of the curvature tensor and its derivatives at $x'$. The leading (quartic) divergence reduces to the vacuum energy in flat space-time and is customarily subtracted from the physical or renormalized value of $\langle T_{\mu \nu} (x') \rangle$ in a general curved space-time. The other state-independent divergences coming from the leading terms in the asymptotic series in (\ref{eq:BP-expansion}), are also subtracted. Alternatively, one can rewrite the full Lagrangian (including the gravitational terms) to include counter-terms having the same form as the expressions involving the Riemann tensor and its derivatives in these UV divergences. Then one can absorb the regularized expressions for these UV infinities into the constants in front of similar terms added to the original Lagrangian (similar to the procedure described above for the i
 nteracting field).  Thus, the renormalization of the free field requires covariant terms involving the curvature tensor to be added to the Lagrangian with coupling constants that are in principle measurable.

Whether one thinks simply of subtracting these state-independent UV-infinite terms from the formal expression for $T_{\mu \nu}$ or thinks of those terms as renormalizing the coupling constants of terms present in the full Lagrangian,  the result is the same, namely, a quantity that has no UV infinities when its expectation value is evaluated for any physical state  in 4-dimensions.  Its expectation value is what we will refer to as the renormalized expectation value of the  energy-momentum tensor, denoted by $\langle T_{\mu \nu} (x') \rangle_{\rm ren}$.\footnote{There are some well known state-independent effects that can be calculated directly from the subtracted terms under discussion, as in the so-called conformal trace anomaly of the energy-momentum tensor.}

Now let us consider the renormalization of the variance $\langle \varphi^2(\vec{x}, t) \rangle$  within the local momentum space method we are considering.
The expression for the renormalized variance $\langle \varphi^2(\vec{x}', t') \rangle_{\rm ren}$ in any given physical state is obtained by subtracting the terms in the asymptotic series that would give UV infinities from the solution for ${\bar G}(k;x')$ in the normal neighborhood of $x'$.    The variance involves no derivatives of $\varphi(x)$, so only the first two terms in Eq.(\ref{eq:BP-expansion}) are subtracted from the local Fourier transform of ${\bar G}(k;x')$. Then the inverse of the local momentum space Fourier transform can be performed, giving a well-defined operator that we will denote by ${\bar G}(x;x')_{\rm ren}$.  The coincidence limit ($x\rightarrow x'$) of the expectation value of this quantity, then gives the renormalized value of the variance at $x'$ in any given physical state.
 Thus, we obtain the result,
\be
\langle \varphi^2(x') \rangle_{\rm ren} = \lim_{x \rightarrow x'}  \langle {\bar G}(x;x')_{\rm ren}  \rangle
\ee
As explained previously, we are subtracting only the minimum number of terms needed to give an expression having no UV divergences.  Even though we found it necessary to subtract four terms of the asymptotic series to obtain $\langle T_{\mu \nu} (x') \rangle_{\rm ren}$, since the renormalization of the variance involves only the first two counter-terms, there is no reason to subtract additional terms in the series. In the gravitational part of the Lagrangian of the renormalized theory the additional covariant terms needed to renormalize the energy-momentum tensor will still appear, but it would be uneconomical and would lead to unnecessary complications to subtract the corresponding momentum space terms from ${\bar G}(k;x')$ in calculating the renormalized variance because those terms are not necessary to regularize UV divergences in the variance.  One does not wish to introduce physical effects into the theory that are not required by the actual renormalization process. Henc
 e, one uses the principle of minimal subtraction. 

In the above discussion of renormalization in a general curved space-time, the asymptotic expansion of ${\bar G}(k;x')$ in (\ref{eq:BP-expansion}) was used to identify the set of UV divergent terms in the local momentum space expansion in the normal neighborhood of $x'$.   The asymptotic expansion was {\em not} used to {\em approximate} a solution of the differential equation for ${\bar G}(k;x')$ that would be the local momentum space Fourier transform of some particular global Green function $G(x, x')$ for values of $x$ in the normal neighborhood of $x'$.
Therefore, employing the {\em asymptotic} series up to fourth (or higher) order to renormalize the variance, expecting that it will give a better approximation to the renormalized value, as unfortunately done in some literature, is simply not correct. Similarly, it is incorrect to argue that the renormalization procedure fails for those values of $k$ and $R$ for which the second term in the expansion (\ref{eq:BP-expansion}) is not small compared to the first term, as claimed in \cite{marozzi11, durrer09}.  This argument would spoil the full renormalization program.
\section{Adiabatic regularization 
in an RW universe}\label{section iii}
A spatially flat RW universe, for which $ds^2 = dt^2 - a(t)^2 (dx^2 + dy^2 + dz^2)$, is often used as the background space-time in discussions of inflation.  In such a universe, the 3-dimensional hypersufaces of constant $t$ are homogeneous and the relevant quantized perturbation field can be expanded as in Eq.(\ref{eq: field in RW}).  The ordinary differential equation satisfied by the mode functions $\varphi_k(t)$ with suitably chosen boundary conditions can be solved exactly for certain forms of the scale factor $a(t)$.  In such cases, one can also obtain the exact expression for the Green function $G(x, x')$ on any given spatial hypersurface.\footnote{If one chooses in an RW universe any point $x'$ on a given hypersurface at $t'$, then the RNC coordinate system with origin at $x'$ will have a normal neighborhood that includes the entire spatial hypersurface at $t'$. However, these RNC coordinates are not the same as the RW coordinate system.}   The method of {\em adiabatic regularization} in the RW universe \cite{parker-fulling74, parker66} starts with the formal expression for $\langle T_{\mu \nu} (x') \rangle$, which is a function of $t'$ in the RW coordinate system.  As described above in connection with the local momentum space method, one can make a large momentum asymptotic expansion of the 3-dimensional Fourier transform of this quantity and identify the leading terms in that expansion that would give UV divergences in the integration over ${\vec k}$.   These terms are the same for any physical state. Now, instead of using dimensional regularization or some other regularization method to find the covariant form of the counter-terms that would be added to the Lagrangian in the course of renormalization, in adiabatic regularization one goes directly to the step of subtracting the relevant leading terms in this momentum space asymptotic expansion from the formal expression for the spatial 3-dimensional Fourier transform of $\langle T_{\mu \nu} (x') \rangle$.  This step insures that the integrand of this 3-dimensional Fourier transform will not give any UV divergences when integrated over 3-momentum.  The renormalized expression for $\langle T_{\mu \nu} (x') \rangle_{\rm ren}$ is defined by this momentum integral.
 
It is worth remarking that just as explained earlier in connection with the $\lambda \phi^4$ interaction, the minimal number of terms in the asymptotic series are subtracted in the process of adiabatic subtraction and renormalization, so as to change the form of the original unrenormalized Lagrangian as little as possible in constructing the renormalized Lagrangian (including the gravitational part of the Lagrangian). 

Note that after making the subtractions in the above method of regularization the integrand of the momentum integral is already regularized (i.e., the momentum space integrand no longer has any UV infinities when it is considered as a whole).  If one wants to study the form of the counter-terms that would appear in the renormalized Lagrangian, then one could dimensionally regularize the individual subtraction terms and see what covariant curvature terms they would involve.  However, in getting the renormalized result in the RW universe, there is no further regularization needed, apart from making the minimum number of subtractions we have described. One of the nice points of this process of adiabatic regularization is that it directly displays the spectral properties of the renormalized physical quantity, such as $\langle T_{\mu \nu} (x') \rangle_{\rm ren}$.

If one is going to subtract the terms in the asymptotic series for all momenta, even small ones, then one has to resolve some ambiguities that are not determined simply by the large k form of those terms.  These possible ambiguities in the asymptotic form of the leading terms (such as whether or not to include the mass $m^2$ along with $k^2$ in the asymptotic series for large $k$), are resolved by requiring that during any intervals of time for which $a(t)$ approaches a constant, the mode functions should approach linear combinations of the positive and negative frequency forms that they would have in Minkowski space. In addition, one can require that in the limit of infinitely slowly changing $a(t)$ the mode functions $\varphi_k(t)$ should take the form of the positive frequency Liouville (or WKB) adiabatic approximation to the mode function solution. This enforces the physically reasonable requirement that the expectation value of the particle number operator is an adiabati
 c invariant; remaining unchanged in the limit of an infinitely slow and smooth expansion of the universe, regardless of the overall change in $a(t)$ during the expansion.

As outlined above for the energy-momentum tensor, the method of {\em adiabatic regularization} of the variance $\langle \varphi^2(x') \rangle$ starts with the formal expression for $\langle \varphi^2(x') \rangle$, which is a function of $t'$ in the RW coordinate system.  As described above in connection with the local momentum space method, one can make a large momentum asymptotic expansion of the 3-dimensional Fourier transform of this quantity and identify the leading terms in that expansion that would give UV divergences in the integration over ${\vec k}$.   These terms are the same for any physical state. In adiabatic regularization one goes directly to the step of subtracting the relevant leading terms in this momentum space asymptotic expansion from the formal expression for the spatial 3-dimensional Fourier transform of $\langle \varphi^2(x') \rangle$.  This step insures that the integrand of this 3-dimensional Fourier transform will not give any UV divergences when in
 tegrated over 3-momentum.  The 
 renormalized expression for $\langle \varphi^2(x')\rangle_{\rm ren}$ is  defined by this momentum integral, regardless of the expansion rate of the universe (assuming the expansion is sufficiently smooth and that infrared divergences that may occur for zero mass are dealt with properly). Therefore, it is again not correct to argue \cite{durrer09, marozzi11} that adiabatic renormalization fails when  the expansion rate $H$ is larger than the physical momentum scale $k/a(t)$, as already pointed out in the previous section. Additionally, in \cite{durrer09} a rather arbitrary redefinition of adiabatic subtraction is made, which does not seem well founded. 

\section {Renormalization in the inflationary universe}

In the physical situation in which the field $\varphi(\vec{x},t)$ represents scalar or (a polarization mode of) tensorial metric perturbation during inflation,   $\varphi(\vec{x},t)$ must be treated as a massless field. The term $V''$ appearing in the wave equation of  scalar perturbations, where $V$ is the potential of a slow-roll inflationary model, should be considered as a second-order adiabatic term \cite{prl09, talk}.
In the massless limit, the subtraction terms defined by the BP renormalization \cite{bunch-parker} and those obtained with the adiabatic renormalization coincide,  thus defining a unique expression for the (renormalized) power spectrum  \cite{prd10}, namely, $\tilde\Delta^2_{\varphi}(k, t)=4\pi k^3(|\varphi_k(t)|^2-C_k(t))$. The subtraction terms $C_k(t)$ can be obtained easily from the two first terms in the expansion (\ref{eq:BP-expansion}) (or equivalently by the adiabatic expansion of the modes).\footnote{In the massless limit the second subtraction term, see (\ref{scalar}) and (\ref{tensor}), has an infrared logarithmic divergence. It can be naturally cured by introducing an infrared cutoff. We shall assume that this low-energy cutoff is far below the relevant cosmological scales.}

For scalar perturbations in single-field, slow-roll inflation one finds (see  \cite{prd10} for details)
\be\label{scalar}
C^{(\delta \phi)}_k(t)=\frac{1}{2(2\pi)^3a^3} \left[\frac{a}{k}+ \frac{a^3}{2k^3}H^2(2-3\eta +5\epsilon)\right] \ ,\ee
where $\delta \phi$ is the inflaton perturbation (in the spatially flat gauge), $a$ is the expansion factor, $H\equiv \dot a/ a$ , and $\epsilon$ and $\eta$ are the usual slow-roll parameters; while for tensorial perturbations one gets
\be \label{tensor}
C^{(h)}_k(t)=\frac{1}{2(2\pi)^3a^3} \left[\frac{a}{k}+ \frac{a^3}{2k^3}H^2(2-\epsilon)\right] ,
\ee
where $h$ stands for one of the two independent fluctuating modes. 
The first term in square brackets in (\ref{scalar}) and (\ref{tensor}) decays rapidly with time during inflation. However, the second term in square brackets varies very slowly during slow-roll inflation and, while maintaining the nearly scale-invariant behavior, it produces a non-negligible effect in the power spectrum even at late times during inflation, as explicitly worked out in \cite{prd10}.

Notice that the physical value of the correlation function, call it  
$\langle \varphi(\vec{x}, t) \varphi(\vec{x}', t) \rangle_{\rm ren}$, at two  separated points $\vec{x}$ and $\vec{x}'$, involves the subtraction terms in such a way as to insure continuity as $\vec{x}$ approaches $\vec{x}'$.  Renormalization can be naturally extended to  higher-order point functions, as for instance the four-point function (odd point-functions are trivially zero)
\bea \langle \varphi(\vec{x}_1,t) \varphi(\vec{x}_2,t) \varphi(\vec{x}_3,t)\varphi(\vec{x}_4,t)\rangle_{\rm ren}&=&  \langle \varphi(\vec{x}_1,t) \varphi(\vec{x}_2,t) \rangle_{\rm ren} \nonumber \\ 
&\times& \langle \varphi(\vec{x}_3,t)\varphi(\vec{x}_4,t)\rangle_{\rm ren}  \nonumber \\ &+& 2 \ \ perm. \ \  . \eea
Therefore, it also maintains   the Gaussianity of  primordial  scalar perturbations. However, the computations beyond leading order in cosmologial  perturbation theory (producing non-Gaussianities)
will be potentially affected by renormalization. Note that the simplest template to generate non-gaussianities \cite{komatsu-spergel} $\varphi(\vec{x}, t) =\varphi_L(\vec{x}, t) + f_{NL}[\varphi_L^2(\vec{x}, t)-\langle \varphi_L^2(\vec{x}, t) \rangle_{\rm ren}]$, where $\varphi_L(\vec{x}, t)$ denotes the linear Gaussian part of the perturbation, assumes implicitly the necessity of renormalizing the variance.

\noindent { \bf Acknowledgements.} This
work has been partially supported by the Spanish grants FIS2008-06078-C03-02, FIS2010-09399-E and NSF grants PHY-0503366
and PHY0854743 and Eberly research funds of Penn State University. We thank A. Ashtekar for stimulating discussions. J. N-S thanks MEC for a sabbatical grant and the   Physics Department of the University of Wiscosin-Milwaukee for their kind hospitality. \\

\end{document}